# The complex superstructure in $Mg_{1-x}Al_xB_2$ at $x \approx 0.5$

H.W. Zandbergen [1,2], M.Y. Wu[1], H. Jiang[1], M.A. Hayward[2], M.K. Haas[2], and R.J. Cava[2]

[1] National Centre for HREM, Laboratory of Materials Science, Delft University of Technology, Rotterdamseweg 137, 2628 AL Delft, The Netherlands

[2] Department of Chemistry and Princeton Materials Institute, Princeton University, Princeton NJ 08544

## Abstract

Electron diffraction and high resolution microscopy have been performed on $Mg_{1-x}Al_xB_2$ with $x \cong 0.5$. This composition displays a superstructure with a repeat period of exactly $2c$ along the $c$ axis and about 10 nm in the $a$-$b$ plane. The superstructure results in ring-shaped superreflections in the diffraction pattern. Irradiation by a strong electron beam results in a loss of the superstructure and a decrease of about 1% in the $c$ lattice parameter. In-situ heating and cooling on the other hand showed that the superstructure is stable from 100 K to 700 K. Possible origins for the superstructure are proposed.

**Introduction**

One of the surprising characteristics of superconducting $MgB_2$ (1) is the simplicity of both its chemistry and structure. The structure consists of honeycomb layers of boron interleaved with simple triangular layers of Mg, and the non-transition metal chemistry appears relatively straightforward. The very limited substitutional chemistry possible (2-4), however, suggests that the structure and chemistry are more complex than they first appear. The substitution of Mg by Al in $Mg_{1-x}Al_xB_2$ further illustrates the complexity (2), The substitution first leads to a decrease in superconducting transition temperature for small x (x < 0.1), and then the disappearance of superconductivity in a manner consistent with expectations from electronic structure calculations (see, for instance, refs. 5-7). However, the loss of superconductivity is accompanied by the partial collapse of the lattice in the direction perpendicular to the boron planes at a chemical miscibility gap $0.1 < x < 0.2$. The compounds with compositions $x > 0.2$ show no bulk superconducting transition down to 4.2K, and observations of superconductivity in that composition range (8) are associated with the difficulty of preparing homogeneous samples. Further complexity was revealed in a study of the $Mg_{1-x}Al_xB_2$ solid solution by electron diffraction (8) which showed the existence of a superlattice near the composition $Mg_{0.5}Al_{0.5}B_2$, apparently indicating a simple doubling of the cell dimension perpendicular to the boron planes at that composition. The cell doubling was attributed to a simple 1:1 ordering of Mg and Al in alternating triangular basal planes.

Here we report a more detailed analysis of the superstructure for $Mg_{0.5}Al_{0.5}B_2$, based on electron diffraction and imaging via high resolution electron microscopy. The study indicates that the superlattice is more complex than first reported, with a significant component of structural modulation within the hexagonal planes as well as perpendicular to the planes. We propose that the interpretation of the superlattice must involve a more complicated picture than simple Mg-Al ordering.

**Experimental**

Samples of stoichiometry $Mg_{1-x}Al_xB_2$ (x= 0.25, 0.45, 0.55, 0.75) were synthesized by direct combination of the elements. Mg flakes (Aldrich Chemical), fine Al powder (Alfa Inorganics) and sub-micron amorphous B powder (Callery Chemical) were mixed in the appropriate stoichiometries and pressed into 0.5g pellets. The pellets were placed on Ta foil, which was in turn placed on a $Al_2O_3$ boat and fired under a mixed 5% $H_2$, 95% Ar atmosphere.

Samples were heated for one 1 hour at 600°C followed by 1 hour at 800°C. After cooling samples were reground and fired for 2 subsequent periods of 2 hours at 800°C with an intermediate regrinding step. An additional set of samples was made from the same starting materials heated in a sealed Ta tube at 1100°C overnight followed by 900 °C for 48 hours. Unless otherwise noted, the results shown are from the series of samples prepared at 800 °C.

Electron microscopy was performed on $Mg_{1-x}Al_xB_2$ samples with compositions x=0.25, 0.45, 0.55 and 0.75. Electron transparent areas were obtained by crushing or ion milling. High resolution electron microscopy (HREM) and electron diffraction (ED) were performed with a Philips CM300UT electron microscope with a field emission gun operated at 300 kV and equipped with Link EDX element analysis equipment. Nanodiffraction was performed using a condenser aperture of 10 μm and an electron probe size of 10-30 nm in diameter. The specimen cooling and heating experiments were performed using Gatan cooling and heating holders and a Philips CM30T electron microscope with a $LaB_6$ gun.

**Results**

Investigation of the samples by EDX in the electron microscope indicated that most of the grains display a rather large variation in the Mg/Al ratio. An example of such a variation is shown in Figure 1. Figure 1a shows a TEM image of a grain of $Mg_{1-x}Al_xB_2$, Element analysis in the areas indicated by A-F shows a variation in the value of x from 0.65 to 0.43, respectively. (Note that a composition variation can also occur along the viewing direction.) In the areas A-F, superstructure reflections are visible, which can – to a first approximation (see below) - be indexed with a unit cell that is doubled along the *c* axis as previously reported (8). The intensities of the superreflections vary for these areas. They are less intense in the areas A-C. Figure 1b, 1c and 1d show the diffraction patterns for areas A, D and F respectively.

The superstructure reflections are strongest for the crystals with compositions close to x = 0.5 in $Mg_{1-x}Al_xB_2$. In the specimen with overall composition x=0.45 the superstructure was observed in all the crystals investigated. Due to the possible loss of Mg during synthesis, the actual Al composition is likely nominally slightly higher than x=0.45, likely Al = 0.5. In the specimen with overall composition x=0.55 only a minority of the crystals showed a superstructure. In the specimens with overall compositions x=0.25 and x=0.75, no superstructures were observed, except for crystal areas with compositions much closer to x=0.5. These results indicate that the appearance of the superstructure is strongly constrained to compositions very close to $Mg_{0.5}Al_{0.5}B_2$.

The superreflections are not single spots exactly in line with the main reflections, but instead consist of double spots. Figure 2 shows three representative electron diffraction patterns. These were taken along the [110], [100] and [3-10] directions of the same area in one crystal. The total specimen tilt from the [110] zone to the [3-10] zone is 52°. The distance between the split spots is 0.19 nm$^{-1}$ for the [110] zone, 0.20 nm$^{-1}$ for the [100] zone and 0.15 nm$^{-1}$ for the [3-10] zone. The distance between the split spots changes a little upon rotation, and it might be that the shape is more like an ellipse or elongated hexagon then a circle. Detailed determination of the shape was not attempted. Comparing the splitting distances of various crystals (and taking only the superreflections close to the origin) some variation is observed, but the variation is generally small. Examples of a normal (0.2 nm$^{-1}$) and the largest splitting (0.35 nm$^{-1}$) are shown in Figure 3.

The splitting of the superstructure reflections is observed continuously when tilting the crystals over 90° about the *c* axis. This implies that the superreflections have approximately a ring-like shape (see figure 4). This ring-like shape is in agreement with the observation that the splitting in the superreflections is the largest near the central spot for crystals exactly aligned along a low index zone (e.g. [100] and [110]). For such crystal orientations, the superreflections at high diffraction angles show no significant split, whereas a gradual increase in the split distance is observed with decreasing diffraction angle. This observation is due to the fact that the curvature of the Ewald sphere results in sampling of the ring at increasing height above the nominal diffraction plane on increasing distance from the origin of the reciprocal space (see Figure 4).

In order to determine whether the observed superstructure is due to a temperature dependent structural modulation, specimens were cooled to 100 K and also heated to temperatures of 500 K, 600 K and 700 K using in-situ TEM specimen cooling/heating holders. Several crystals were first selected at room temperature. Then the specimen was either cooled or heated and the same crystals were investigated at different temperatures. In all cases the superstructure was present and its appearance was unchanged. The superstructure is therefore not associated with a thermally induced structural change in the temperature region between 100K and 700K.

Although the superstructure cannot be removed by in situ cooling or heating (up to 700K), irradiation of the specimen by the electron beam leads to a loss of the superstructure (figure 5). This rate at which the superstructure disappears depends on the electron dose and varies from crystal to crystal. This disappearance takes at least several minutes with normal HREM imaging (10$^6$ electrons per nm$^2$ per second), but can occur within several seconds when a more intense beam is used. The right part of the crystal area shown in this figure has been irradiated for 1 minute with an electron beam of a diameter of about 30 nm, 10 times stronger than used for normal HREM imaging, resulting in the local loss of the superstructure. This suggests that the disappearance of the superstructure under irradiation is associated with a process which changes the composition of the material in the electron beam. Analysis of the Fourier transforms of the areas in Figure 5 showing the presence and absence of the superstructure indicates that in the area where the superlattice has been lost, the $c$ axis has contracted by approximately 1%. Before the irradiation the $c$ axes of both areas were equal.

The change in the superstructure by the electron beam is probably initiated or caused by the knockout of some atoms. Thus the irradiation is likely to result in a small change in composition. The reduction in the length of the $c$ axis of 1% in this process is quite large. If due to a change in the Mg/Al ratio, for instance, it would be consistent with a loss in Mg equivalent to a change in x from x=0.45 to about c=0.60 (assuming a linear dependence of the $c$ axis on the Al/Mg ratio). However, such a compositional change is highly unlikely because Mg and Al have a similar knockout probability, and no substituting Al atoms are available. Thus a relatively small removal of Mg, Al, or in particular B from the lattice, resulting in point defects, must result in this lattice contraction.

Figure 6 is a [210] HREM image of $Mg_{1-x}Al_xB_2$ with x=0.45 showing the superstructure in real space. This direction is employed because the low index orientations [100] and [110] show a less dominant superstructure. (This is not related to the intensities of the superreflections in these orientations but to the much stronger main reflections.) The doubling of the $c$ axis of the simple $MgB_2$ structure is clearly observed. The Fourier transform of this HREM image is shown in the inset. The Fourier transform manifests the splitting of the superreflections seen in the electron diffraction patterns. In the HREM image, this splitting can be observed as a faint variation in the contrast with a spacing of about 12 nm in the $ab$ plane and $2c$ along the $c$ axis. Figure 7 shows an HREM image of $Mg_{1-x}Al_xB_2$ (x = 0.45) in the [100] orientation. The superstructure is faintly visible along both the $c$ axis and the $b$ axis. The Fourier transform is shown as an inset. Due to the dominant lattice image of the average structure, the superstructure is only weakly visible. For this crystal the superstructure reflections are less dominant, also because it has already been partly destroyed.

The strongest superreflection intensities are observed for the compositions close to x = 0.5, as previously mentioned. The superstructure intensities were also found to depend on the synthesis temperature: they were much more visible in the specimens annealed at 800°C than in those annealed at 900°C. The reason for this is not known at this time. Further investigation would be required to determine the origin for this difference if it proved to be of interest.

The split superreflections can be indexed with a fourth index, $m$, as is done routinely in four-dimensional spacegroups. If the basic the hexagonal unit cell is taken as $a$=0.30 nm and $c$=0.67 nm, e.g. with the normal $c$ axis doubled, then only superreflections with m=1 are visible as satellites of the *hkl* reflections with $l$=2n+1. m=2 reflections might potentially be present (but not detectable) in the shoulders of the main reflections (*hkl* reflections with $l$=2n), but they are certainly not visible as satellites of the *hkl* reflections with $l$=2n+1. The presence of only the first order superreflections indicates that the superstructure can be considered as only a sinusoidal modulation of the basic structure. This means that only a gradual structural modulation has caused the superstructure: there are no higher order harmonics than the first sine term involved. Since the superstructure yields diffraction rings, the superstructure ordering must vary from area to area. The superlattice modulation, if circle-like, may alternately be looked at in a different way, taking the basic $MgB_2$ unit cell ($a$=0.30 nm and $c$=0.35 nm) as a basis. In this basis, the superlattice reflections indicate the presence of a sinusoidal structural modulation in $Mg_{0.5}Al_{0.5}B_2$ represented in reciprocal space by the wave vector $q = (q_1, q_2, 0.5)$, where to a first approximation $(q_1^2 + q_2^2)^{1/2} \approx 0.1$: a structural modulation of the basic $MgB_2$ structure which doubles the simple $c$ axis but also contains a significant component in the hexagonal $a$-$b$ plane.

**Discussion**

Several causes can be proposed for the observed superstructure. Firstly, as originally proposed (8), ordering of Al and Mg can cause a superstructure. Given the hexagonal unit cells of $MgB_2$ ($a$=0.3083 nm, $c$=0.3521 nm) and $AlB_2$ ($a$=0.300 nm, $c$=0.3245 nm) and the sizes of metallic Mg (0.160 nm) and Al (0.143 nm), the most logical Al-Mg ordering is a near neighbor alternation in three directions. This would result in a rather simple superstructure. Two models for such ordering are given in Figure 9a and 9b. The model shown in Figure 8a, with simple alternation of Al and Mg along the $c$ axis as suggested in ref 8, would yield approximately the observed diffraction patterns, but not the splitting of the superlattice peaks. If an Al/Mg ordering is the cause of the superstructure, the observed split in the superreflections would require a long-range distribution of the Mg and Al atoms within the hexagonal planes. The presence of an ordering of patches of about 20 by 20 Al atoms and patches of about 20 by 20 Mg atoms, for instance, would account for the observed diffraction patterns. Several arguments can be proposed against an ordering of the Al and Mg as being the cause of the superstructure. Although the 1:1 ratio is often a special ordering composition, one would expect other types of superstructure periods along the *c* axis at other Mg-Al ratios if such ordering were energetically favorable. This is not observed, however. Secondly, the loss of superstructure by the electron beam and not by in situ heating at 400°C suggests that the loss of superstructure is due to a change in composition rather than a rearrangement of the Al and Mg atoms. The disappearance of the superstructure seen in Figure 5 would require a complete reshuffling of the Al and Mg atoms over relatively long distances in the hexagonal layers if that were the origin of the superstructure, which is highly unlikely.

The second possible cause for the superstructure is a rearrangement of the B atoms due to B excess or deficiency. One possible modulation resulting from the insertion of extra B atoms that would successfully describe the superstructure is shown in figure 8d. If the superstructure is caused by boron nonstoichiometry which accompanies Al substitution in $MgB_2$, then the excess (or deficient) boron would be expected to enter the lattice in a continuous fashion with a continuously changing Al content. The structural modulation would then be observed over a wide range of Al contents and its wave vector would change continuously with Al content. That is not observed, however, and it is difficult to reconcile the especially strong superlattice near x=0.5 and absence of a superlattice at high and low Al contents with such a model.

The third possible cause for the superstructure is the presence of a structural instability associated to the changes in the unit cell due to the changing Al/Mg ratio. $AlB_2$ and $MgB_2$ have rather different *c* axis parameters though their in-plane dimensions are similar ($MgB_2$, *a*=0.3083 nm, *c*=0.3521 nm, and $AlB_2$, *a*=0.300 nm, *c*=0.3245 nm). The miscibility gap in the composition region 0.1<x<0.2 (2) is a direct indication that certain *c* axes at certain compositions are less stable. Furthermore, the *a* axis is relatively small in both $MgB_2$ and $AlB_2$ and close to the boundary beyond which the B planes are not flat but buckled. It might be that the boundary between a buckled B layer and a non-buckled layer also depends on the *c* parameter. If this is the case, then a decrease in the *c* axis due to a substitution of Mg by Al might trigger a partial bucking of the B planes, resulting in a superstructure. To explain the superstructure one needs an alternation of buckled and non-buckled B layers along the *c* axis, and also an alternation of buckled and non-buckled planes in the *a-b* plane. The disappearance of the superstructure $Mg_{1-x}Al_xB_2$ with x≈0.5 by irradiation would then be caused by the change in the lattice parameters on inducing a change in composition. An important argument against this model is that the actual a axis for $Mg_{1-x}Al_xB_2$ with x≈0.5 is 0.305 nm, which is relatively far from the actual stability boundary for non-buckled B planes.

Finally, the superlattice may be due to an electronic instability at the electron count near 0.5 extra electrons per $MgB_2$ formula unit. If, for example, there were a particularly favorable case of Fermi surface nesting at that electron count, then a structural distortion is expected to occur with the same structural wavevector as the electronic wavevector which describes the nesting, as commonly observed in materials which display charge density waves. If the superstructure is indeed due to an electronic instability of this type, then there may be implications for the mechanism of superconductivity in $MgB_2$. Further theoretical work, however, will be necessary to explore that possibility.

The present work has reported the characteristics of the complex superstructure occurring in the $Mg_{1-x}Al_xB_2$ layered phase near the composition x = 0.5 in both real space and reciprocal space. Its complexity strongly reinforces the idea that $MgB_2$, in spite of its apparent structural and chemical simplicity, is a very special and unusual material from a crystal-chemical viewpoint. Though the origin of the superlattice is not known at this time, several of the possible origins have implications for the occurrence of superconductivity and are worthy of further investigation.

**Acknowledgement**

The work at Delft was supported by the Nederlandse Stichting voor Fundamenteel Onderzoek der Materie (FOM). The work at Princeton was supported by the US Department of Energy grant DE-FG02-98-ER45706 and the NSF MRSEC program, grant DMR9808941.

**Figure captions**

Figure 1.
a) TEM image of a grain of $Mg_{1-x}Al_xB_2$. from a specimen with nominal composition x=0.45. Electron diffraction and EDX element analysis has been done on the areas A-F. The EDX results are given as inset in a). b) - d) show the electron diffraction patterns of the areas A, D and F respectively. Note that the relative intensity of the superreflections in A in much less than in c) and d).

Figure 2.
Electron diffraction patterns of the same area along a) [110], b [100] and c) [3-10]. Superstructure reflections are observed in each of these three orientations as well as in the crystal orientations between these directions.

Figure 3.
Electron diffraction patterns with a rather small and a large spacing between the pairs of superreflections.

Figure 4.
Schematic representation off the 00*l* line in diffraction space with the ring-shaped superreflections. The intersections of the reflections with the Ewald sphere are given by black dots. Due to the curvature of the Ewald sphere the intersections of the Ewald sphere with the rings results in pairs of spots, of which the spacing changes from the maximum spacing near the center to a coincidence when the Ewald sphere is just touching the diffraction ring.

Figure 5.
TEM image showing the effect of irradiation with the electron beam. A beam with an intensity of about $10^7$ electrons per second $nm^2$ has been positioned for one minute on the right side of the crystal shown in this figure (indicated by the circle). Due to this irradiation the superstructure has almost disappeared. The inset shows the local superstructure ordering determined from the intensity of the 001 superstructure reflection in the Fourier transform of a 128 by 128 pixels window that was scanned over the image.

Figure 6.

[3-10] HREM image showing the superstructure. On the left a part of the Fourier transform showing in particular the split of the superreflections is given.

Figure 7.

[100] HREM image showing a faint superstructure. A relevant part of the Fourier transform of part of this image is shown as inset.

Figure 8

Models for the superstructure in $Mg_{1-x}Al_xB_2$. Only the Mg and Al atoms are depicted. a) gives an ordering of the Mg and Al atoms with alternating planes of Al and Mg along the *c* axis. This results in a $a=a_0$, $c=2c_0$ superstructure. b) gives an ordering with as many as possible Mg atoms around an Al atom and vice versa. This leads to an I centered orthorhombic unit cell with $a=a_0$, $b=a_0\sqrt{3}$, $c=2c_0$. c) and d) show models of an ordering of Al and Mg and a change in the separations between the (Mg,Al) planes due to a change in the B plane, respectively, which both would give rise to the observed superstructure. Note that for the sake of a rather simple model the actual number of Mg and Al atoms along the in-plane modulation direction is reduced (12 in stead of about 40). Because of the sinusoidal character of the modulation the changes in the local atomic arrangement should be only gradual. The unit cell is indicated with a black rectangle. a'), b'), c') and d') give the diffraction patterns for the structure models of a), b), c) and d).

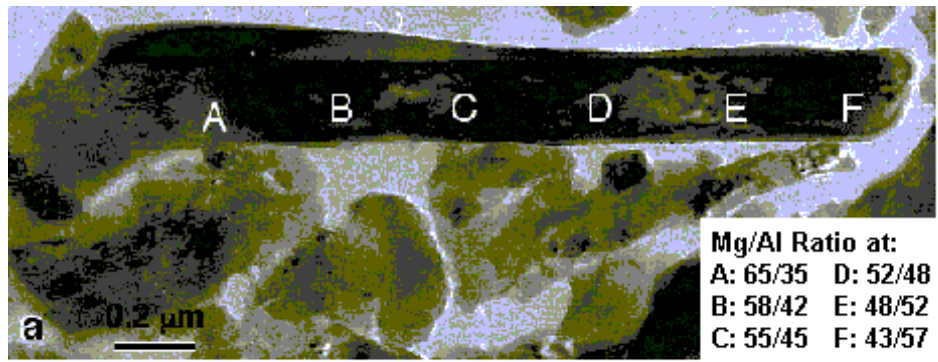

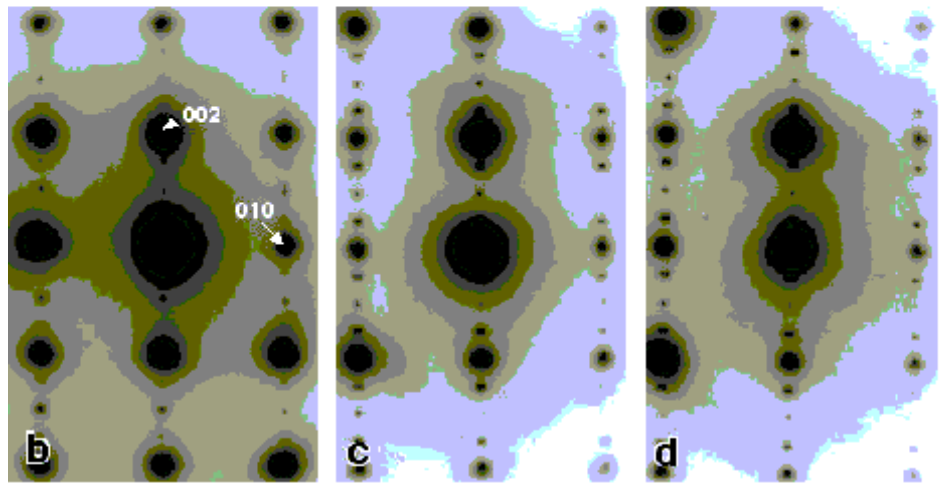

Figure 1

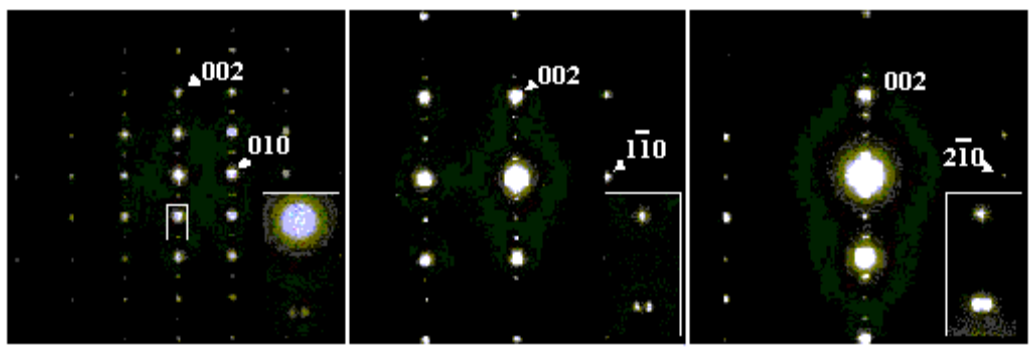

Figure 2

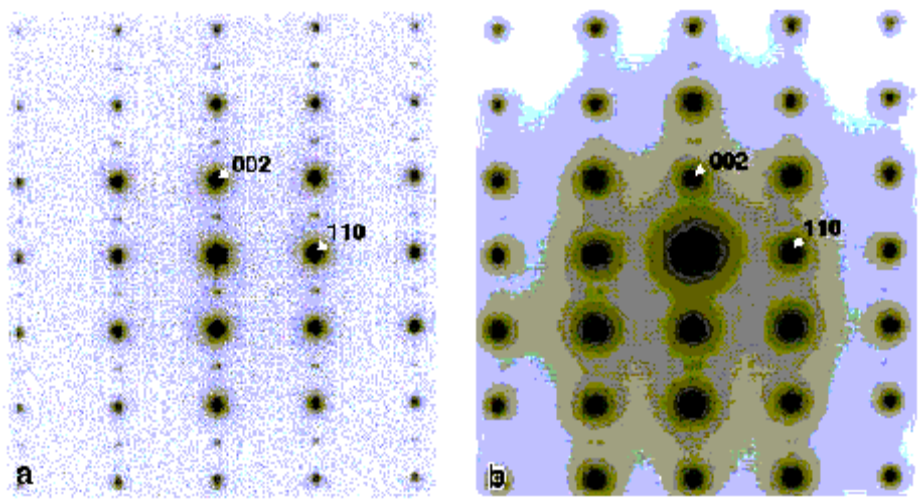

**Figure 3**

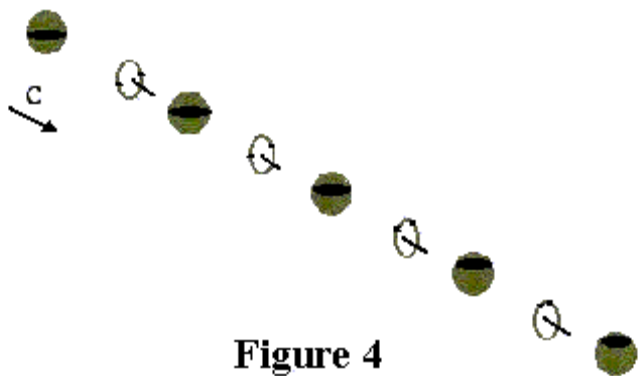

**Figure 4**

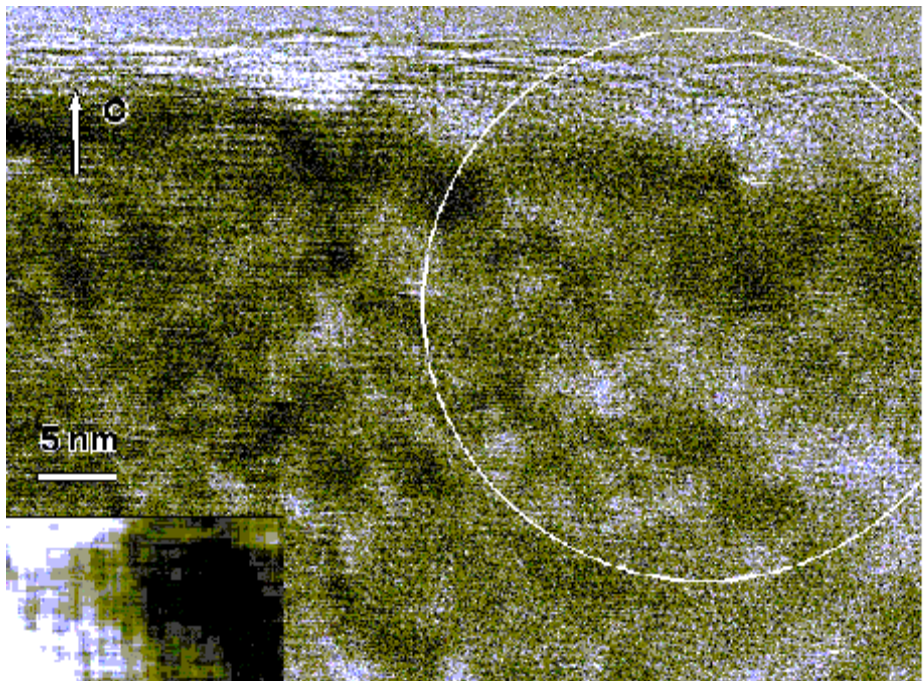

Figure 5

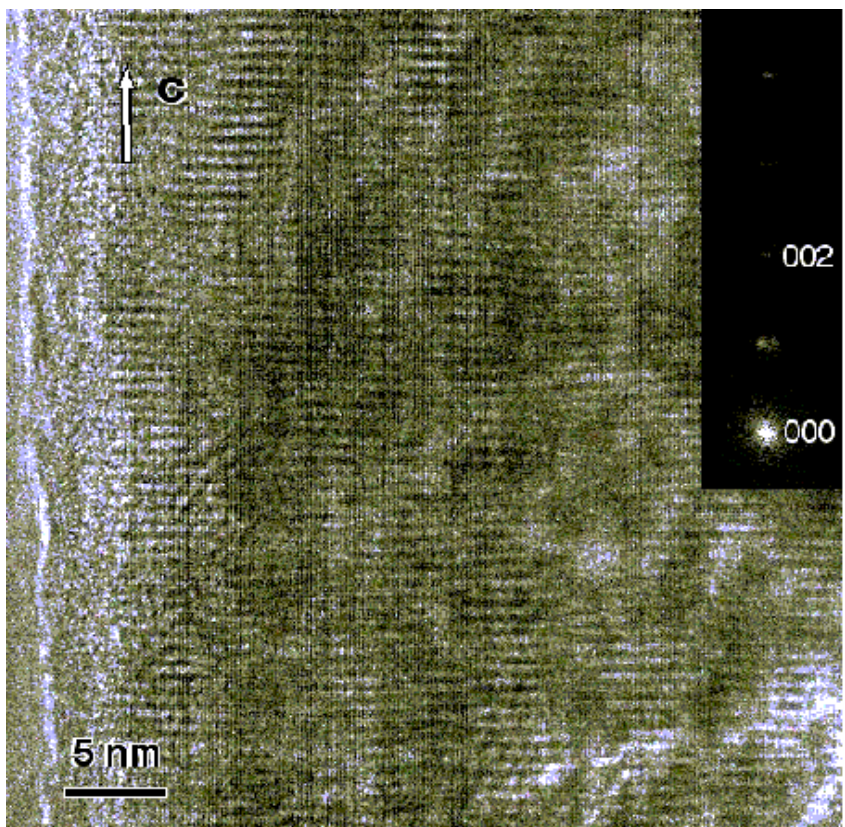

Figure 6

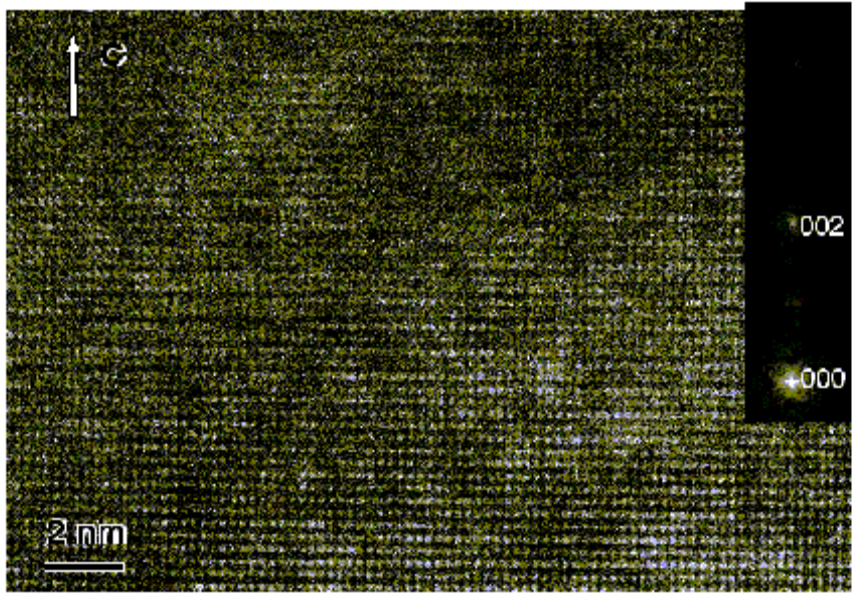

Figure 7

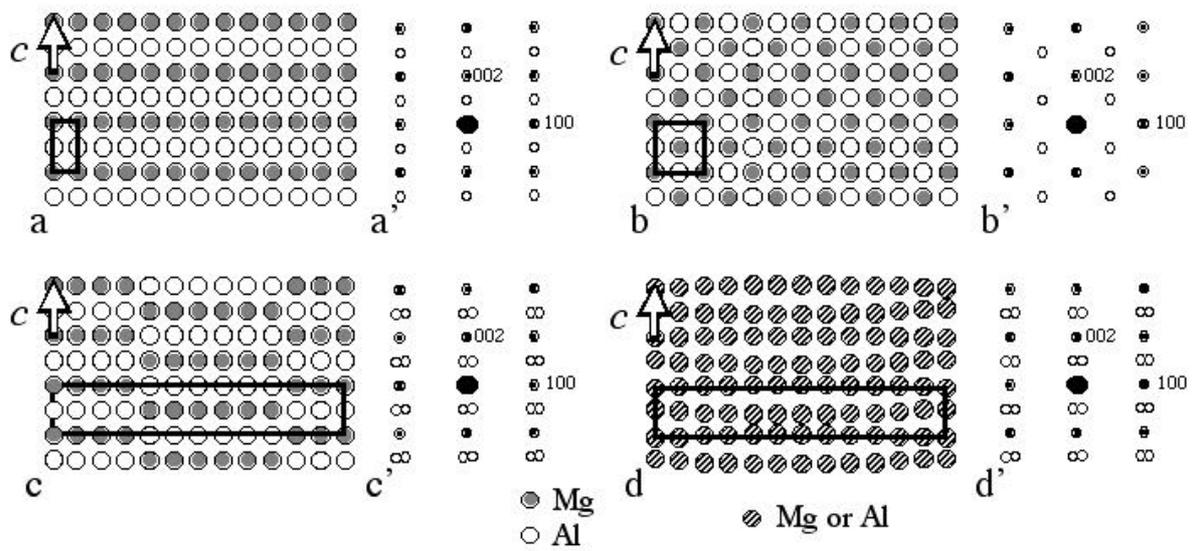

Fig. 8